\documentclass[5p]{elsarticle}
\usepackage{hyperref}
\usepackage{xcolor}
\usepackage{amsmath}
\usepackage{amsfonts}
\usepackage{amssymb}
\usepackage{graphicx}
\usepackage{subfigure}
\journal{}

\begin{document}
\begin{frontmatter}
\title{Search for high energy $\gamma$-rays \\ from the direction of the candidate electromagnetic counterpart \\ to the binary black hole merger
gravitational-wave event S190521g}
\author[Scobeltsyn,MSU]{Egor Podlesnyi\corref{correspondingauthor1}}
\cortext[correspondingauthor1]{Corresponding author}
\ead{podlesnyi.ei14@physics.msu.ru}
\author[Scobeltsyn,ICRR]{Timur Dzhatdoev\corref{correspondingauthor2}}
\cortext[correspondingauthor2]{Corresponding author}
\ead{timur1606@gmail.com}

\address[Scobeltsyn]{Federal State Budget Educational Institution of Higher Education, M.V. Lomonosov Moscow State University, Skobeltsyn Institute of Nuclear Physics (SINP MSU), 1(2), Leninskie gory, GSP-1, 119991 Moscow, Russia}
\address[MSU]{Federal State Budget Educational Institution of Higher Education, M.V. Lomonosov Moscow State University, Department of Physics, 1(2), Leninskie gory, GSP-1, 119991 Moscow, Russia}
\address[ICRR]{Institute for Cosmic Ray Research, University of Tokyo, 5-1-5 Kashiwanoha, Kashiwa, Japan}

\begin{abstract}
The gravitational-wave event S190521g --- a likely binary black hole merger in the accretion disk of an active galactic nucleus --- was accompanied by an optical counterpart. Such dense environments around luminous energy release regions are favourable for high energy $\gamma$-ray production. We report on a search for high energy $\gamma$-rays from the direction of the candidate electromagnetic counterpart to the S190521g event using publicly-available data of the {\it Fermi-LAT} space $\gamma$-ray telescope. No significant signal was found. We present upper limits on the spectral energy distribution of the source in the 100 MeV -- 300 GeV energy range. We discuss the importance of studying S190521g-like transients in the context of cosmic ray acceleration, $\gamma$-ray and neutrino production in such sources.
\end{abstract}
\end{frontmatter}

\section{Introduction}

Very recently, the first detection of a plausible optical electromagnetic counterpart to a candidate binary black hole merger S190521g was reported in \cite{Graham2020}. Namely, an optical flare with the duration of $t_{opt} \sim 50$ days was detected with the {\it Zwicky Transient Facility}, indicating that this merger occured inside the accretion disk of J124942.3 + 344929\footnote{hereafter called J1249 + 3449 for simplicity}~--- an active galactic nucleus situated at the redshift of $z = 0.438$.

This observation is interesting in the context of \mbox{$\gamma$-ray} astronomy for the following two reasons: 1)~a~very high value of the estimated bolometric luminosity of the flare $L_{bol} \sim 10^{45}$~erg/s\footnote{given that the estimated mass of the final black hole is $\sim 100$ solar masses \cite{Graham2020}}, 2) a relatively dense environment around the energy release region --- both in terms of matter\footnote{the estimated gas density in the accretion disk is $\sim 10^{-10}$ g/cm$^{3}$~\cite{Graham2020}} and background photon density. Indeed, the Eddington ratio for J1249 + 3449 is $\sim$0.02-0.2, typical for quasars (e.g. \cite{Padovani2019}) that usually have dense accretion flows \cite{Shakura1973} as well as broad line regions (BLRs), filled with photon fields reflected from BLR clouds. These circumstances make the region around the merger favourable for particle acceleration and subsequent $\gamma$-ray and neutrino production. 

For the apparent isotropic bolometric luminosity $L_{bol} \sim 10^{45}$~erg/s $\sim 10^{57}$ eV/s and the luminosity distance to the source $d_{L} \sim 3\cdot 10^{3}$ Mpc $\sim 10^{28}$ cm the expected bolometric energy flux could be as high as $F_{E}= L_{bol}/(4 \pi d_{L}^{2}) \sim 1$ eV/(cm$^{2}$s). If the power transferred to high energy (HE, $E > 100$ MeV) particles is comparable with $L_{bol}$\footnote{some astrophysical objects exhibit the peak in the observable spectral energy distribution beyond 10 TeV (e.g. \cite{Biteau2020}), indicating that such ``extreme accelerators'' do indeed exist}, the HE $\gamma$-ray flux (from 100 MeV to 300 GeV) $F_{\gamma} \sim F_{E}$.\footnote{neglecting cosmological effects} $F_{\gamma} \sim 1$ eV/(cm$^{2}$s) is still within the capabilities of the {\it Fermi-LAT} space $\gamma$-ray telescope \cite{Atwood2009}, motivating the search for HE $\gamma$-rays from this source.

Assuming the diffusive shock acceleration mechanism, the characteristic acceleration timescale is $t_{acc}= cE/(3eBv_{s}^{2})$ (e.g. \cite{Malkov2001,Uchiyama2007}, where $c$ is speed of light, $e$ --- the elementary charge, $E$ --- the energy, $B$ --- the magnetic field, and $v_{s}$ --- the shock front velocity. For $E= 10$ GeV, $B = 1$ G \cite{Silantev2009}, and $v_{s} = 200$ km/s $t_{acc} \sim 10^{3}$ s $\ll t_{opt}$. Therefore, the HE $\gamma$-ray signal of a leptonic nature is expected to be timely coincident with the optical flare. For the photohadronic emission mechanism, however, the expected value of $t_{acc} \sim 10^{7}$ s even for $B = 10$ G due to the high energy threshold of this mechanism ($\sim 10^{15}$ eV).

In the present paper we perform a search for HE \mbox{$\gamma$-rays} from the direction of J1249 + 3449 on a month -- year timescale using publicly-available data of the {\it Fermi-LAT} space $\gamma$-ray telescope. We note that no significant signal was found with {\it Fermi-LAT} over a short time period of 10~ks \cite{Axelsson2019}.

\section{{\it Fermi-LAT} data analysis \label{sect:analysis}}

Here we derive upper limits on the observable spectral energy distribution (SED = $E^{2}dN/dE$) of J1249 + 3449. We select \textit{Fermi-LAT} data within two time windows: 1) 2019, May 19, 00:00:01 UTC --- 2019, September 1, 00:00:01 UTC; 2) 2019, May 19, 00:00:01 UTC --- 2020, June 30, 00:00:01 UTC. The region of interest (ROI) is a circle with the radius of $15^{\circ}$, centered at the position of J1249 + 3449 ($\alpha_{\mathrm{J2000}} = 192.426^{\circ}, \delta_{\mathrm{J2000}} = +34.8247^{\circ}$) which was associated with the gravitational-wave event. We have applied the energy selection from $100~\mathrm{MeV}$ to $300~\mathrm{GeV}$ and used standard recommendations for off-plane point source identification with \textit{Fermi-LAT}.

\begin{figure}[tb]
\centering
\includegraphics[width=0.53\textwidth]{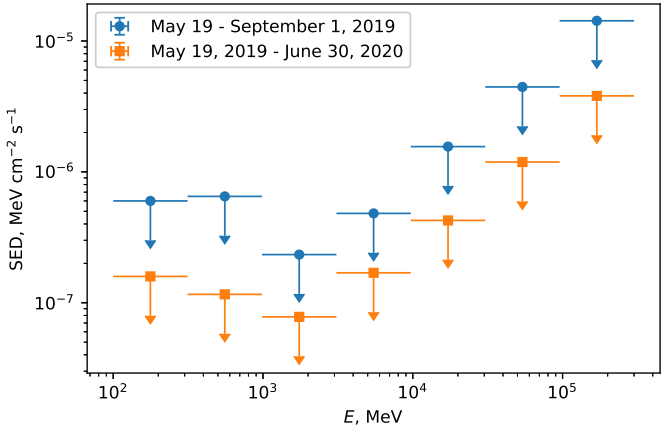}\hspace{2pc}%
\begin{minipage}[t]{16pc}\caption{Upper limits on the SED of J1249 + 3449 over the time period of 105 days from 19$^{th}$ of May, 2019 till 1$^{st}$ of September, 2019 and over the time period of 408 days from 19$^{th}$ of May, 2019 till 30$^{th}$ of June, 2020.}
\label{Fig1}
\end{minipage}
\end{figure}

We performed an unbinned likelihood analysis of \texttt{P8R3 SOURCE} class events with \texttt{P8R3\_SOURCE\_V2} instrument response function\footnote{\url{https://fermi.gsfc.nasa.gov/ssc/data/analysis/documentation/Cicerone/Cicerone_LAT_IRFs/IRF_overview.html}}. We constructed a model of the observed emission including the following sources that could contribute to the detected $\gamma$-ray counts inside the ROI: 1) J1249 + 3449 itself, modelled as a point-like source with a power-law spectrum, situated at the center of the ROI, 2) all sources from the Fermi 8-Year Point Source Catalog (4FGL) \cite{Abdollahi2020} located within $20^{\circ}$ from the center of the ROI, 3) the galactic $\gamma$-ray background according to the model $\texttt{gll\_iem\_v07}$ and 4) the isotropic diffuse $\gamma$-ray background according to the model $\texttt{iso\_P8R3\_SOURCE\_V2\_v1}$\footnote{\url{https://fermi.gsfc.nasa.gov/ssc/data/access/lat/BackgroundModels.html}}. For J1249 + 3449 we set both spectral index and normalization as free parameters; for point-like sources within $5^{\circ}$ from the center of the ROI and the diffuse backgrounds only the normalizations were left free, while the spectral shapes were fixed; for point-like sources beyond $5^{\circ}$ from the center of the ROI both normalizations and shapes were fixed according to the catalog values.

Using this model of the observed emission, we perform the maximization of the likelihood. We calculate the value of the test statistic $TS$ corresponding to the hypothesis of the J1249 + 3449 emission being present in the dataset against the null hypothesis of it being absent. We obtain $TS \ll 1$, i.e. no significant $\gamma$-ray flux was detected from this object. Then we derive upper limits (95\% confidence level) on the SED, using \textit{likeSED.py} \cite{JohnsonFSSC}. These upper limits for both considered time intervals are shown in Fig.~\ref{Fig1}. We also performed an independent binned analysis using the \textit{fermipy} package \cite{Wood2017} and derived upper limits with the \textit{gta.sed} method implemented in this package, but the obtained limits are weaker than for the case of the unbinned analysis.

\section{Discussion \label{sect:discussion}}

There are many possible explanations for the negative results of the search for HE $\gamma$-rays reported above, including the following ones:\\
1) the bolometric luminosity $L_{bol}$ and/or the power transferred to HE particles could have been significantly overestimated;\\
2) HE $\gamma$-ray production efficiency could be significantly lower than unity (especially for the case of primary protons or nuclei);\\
3) HE $\gamma$-rays could have been beamed out away from the line-of-sight;\\
4) HE $\gamma$-rays could have been absorbed by the material of the accretion disk, or photon fields of the accretion disk corona and/or thermal photon field created by hot gas around the merger. The model curves presented in \cite{Costamante2018} demonstrate that the absorption of $\gamma$-rays on photon fields of the BLR is usually significant at $E > 10-30$ GeV. This might impair the prospects of detecting S190521g-like transients with imaging atmospheric Cherenkov telescopes such as H.E.S.S. \cite{Hinton2004,Bonnefoy2018}, MAGIC \cite{Aleksic2016a,Aleksic2016b}, VERITAS \cite{Krennrich2004,Park2016}, or CTA \cite{Actis2011,Acharya2013}.

A detailed study of these effects is underway and will be published elsewhere. X-ray data may be helpful in constraining some models, especially those that include the process of electromagnetic cascade development in the source (both in matter and photon fields) with subsequent synchrotron emission of cascade electrons.

Of course, other multiwavelenght/multimessenger data could also be helpful, in particular, very high energy (VHE, $E > 100$ GeV) neutrino searches from the direction of J1249 + 3449. S190521g-like transients could be copious\vspace{1cm} sources of two components of VHE neutrinos: 1) ``hadronuclear'' neutrinos coming from interactions of accelerated protons or nuclei with the material of the accretion disk and 2) ``photohadronic'' neutrinos from interactions of these protons or nuclei with photon fields in the source. Such neutrinos could contribute to the IceCube diffuse neutrino flux \cite{Aartsen2013a,Aartsen2013b}. A characteristic signature of this two-component neutrino flux is an ``ankle'' connecting relatively hard photohadronic component and a softer hadronuclear component.

Future observations would likely result in the identification of more S190521g-like events. A stacking analysis of such sources using {\it Fermi-LAT} data could result in the detection of a significant HE $\gamma$-ray signal. In this case, it would become possible to constrain the efficiency of the particle acceleration mechanism and the parameters of the optical-UV photon fields in the source using the effect of pair production by two photons $\gamma\gamma\rightarrow e^{+}e^{-}$ (in particular, the far-UV photon fields that could not be directly observed). This, in turn, would make it possible to constrain the models of $\gamma$-ray, cosmic-ray, and neutrino production in S190521g-like transients.

\section*{Acknowledgements}

The reported study was funded by RFBR, project number 20-32-70169. We are grateful to the organizers of the researcher school ``Multimessenger data analysis in the era of CTA'' (Sexten, Italy, 2019) for tutorials provided by them. E. I. Podlesnyi thanks the Foundation for the Advancement of Theoretical Physics and Mathematics ``BASIS'' for the support in participation at the aforementioned school (travel-grant no. 19-28-030) and for the student scholarship (agreement no. 19-2-6-195-1).

\section*{References}
\bibliography{GW-Gamma.bib}
\end{document}